\documentclass[11pt]{article} 

\usepackage[utf8]{inputenc} 

\usepackage[T1]{fontenc}
\usepackage{currvita}
\usepackage[noblocks]{authblk}     

\usepackage{graphicx}
\usepackage{caption}
\usepackage{subcaption}

\usepackage{geometry} 
\usepackage{booktabs} 
\usepackage{array} 
\usepackage{paralist} 
\usepackage{verbatim} 
\usepackage{subfig} 
\usepackage{url}

\usepackage{fancyhdr} 
\usepackage{sectsty}
\allsectionsfont{\sffamily\mdseries\upshape} 

\usepackage[nottoc,notlof,notlot]{tocbibind} 
\usepackage[titles,subfigure]{tocloft} 

\title{A case study of conspiracy theories\\ about Charlie Hebdo terrorist attack}
\author{Nata\v{s}a Golo\\ Cevipof, Center for political research, SciencesPo, Paris, France\\ Email: natasa.golo@gmail.com}

\begin{document}
\maketitle

\begin{abstract}
The results of the public opinion poll performed in January 2015, just after the terrorist attack on the French satirical weekly magazine Charlie Hebdo and the kosher supermarket in Paris, when 17 people were killed, showed that a significant number of French citizens held conspiratorial beliefs about it $(17 \%)$. This gave reason to an alternative analysis of public opinion, presented in this paper. We collected 990 on-line articles mentioning Charlie Hebdo from Le Monde web site (one of the leading French news agencies), and looked at the ones that contained words related with conspiracy (in French: `complot', `conspiration' or `conjuration'). Then we analyzed the readers response,  performing a semantic analysis of the 16490 comments posted on-line as reaction to the above articles. We identified 2 attempts to launch a conspiratorial rumour. A more recent Le Monde article, which reflects on those early conspiratorial attempts from a rational perspective, and the commentary thereon, showed that the readers have more interest in understanding the possible causes for the onset of conspiratorial beliefs then to delve into the arguments that the conspiracists previously brought up to the public. We discuss the results of the above semantic analysis and give interpretation of the opinion dynamics measured in the data.
\end{abstract}


\section{Introduction}
\label{sec:1}

Since the 9/11 attack, the intensity of counter-terrorism measures, as well as research, has significantly increased.
Nevertheless, fear management as a way of counter-terrorism is still a discipline to be developed \cite{Bakker2014}.
According to Bakker et al. \cite{Bakker2014},  fear management has to do with all the efforts of governmental institutions, prior, during and after a terrorist attack or treat, to manipulate the human capital in society in order to improve the
 resilience of that society. They define fear as a sentiment of anxiety caused by the perception of danger and tie them
to the analytical category of coping mechanisms, as developed in psychological and social scientist research.

On another hand, crisis communication after terrorist attacks and strengthening public resilience
against terrorism are among the fifty un- and under researched topics in the field of (counter-) terrorism studies,
as identified by Schmid \cite{Schmidt2011}. One example on that list, No. 18, relates to the link between the media, the internet and terrorism, questioning whether they impact each other, and what can be done to control
them while upholding the freedom of speech, the freedom of expression.

One way to sense the level of anxiety in the society could be to monitor the level of penetration of conspiratorial beliefs. In the past, conspiracy theories used to be defined and implemented by a small number of Machiavellian individuals. However, with the current speed of communication they might get much larger dimension and activate complex and unpredictable social dynamics. Mechanisms extending the so called "wisdom of the crowd" emergence, may work backwards and transform a  poorly informed, half-wit, relatively inert, disconnected population into a well integrated, deadly efficient machine of misinformation, physical and cultural destruction.   
In the context of terrorism, conspiracy theories might become an additional tool to build up gradually the arguments serving to the terrorists goals. In the sense of a self-fulfilling prophecy, a conspiracy theory can recruit the agents that will serve to the goals of the conspiracists in the future. In fact it can lead to some of the acolytes taking or supporting actions which in themselves are real conspiracies.  

In terms of minority opinion spreading, conspiratorial beliefs were previously analysed in \cite{bib1}, \cite{bib2}. A model of rumours of a 9/11 related hoax is proposed in \cite{bib3}. It is however very difficult to provide the empirical support for the dynamics / propagation of conspiratorial beliefs. It was argued that the eventual percolation of the groups of acolytes into a macroscopic network is a crucial factor in allowing terrorist operatives to move and act in otherwise "foreign territory".
Thus the identification and characterization of the overt activity involved in generating, propagating and supporting conspiracy theories is part of the wider goal of understanding, modelling and influencing the dynamics of subversive groups, organizations and their mass social groups support.

Studies indicate that this chain reaction might onset at any time after the terrorist attack, and the dynamics after the onset might span over a period of several years \cite{Golo2015}. Bessi et al. \cite{Bessi2015} measured how information related to a conspiracy news is consumed on Facebook. They tested potential biases induced by the continued exposure to unsubstantiated rumours on users' content selection, and concluded that 77.92 \% of likes and 80.86 \% of comments are from users usually interacting with conspiracy stories. This is congenial with the previous finding of the authors \cite{Golo2015}, who found that the probability to have a strong opinion for or against a conspiracy theory grows with the users activity.

\subsection{Background information about the terrorist attack}

The 2015 attack to Charlie Hebdo office in Paris roots in the conflict of beliefs between the religious radicalists who objected to the use of the caricature of the prophet Mohamed and the journalists who defended the freedom of speech in terms of religion, which is guaranteed by the French law (``La{i}cit\'{e}'', French term for secularism, is currently a core concept in the French constitution).

Here is a short description of the events. On the $7^{th}$ January 2015, two Islamist gunmen forced their way into and opened fire in the Paris headquarters of Charlie Hebdo, killing twelve. Simultaneously a second attach happened in a kosher grocery store in Paris, killing 5.  As a result, the French government organized an event when several million people marched across France, lead by 50 world leaders in Paris in tribute for the terror victims ('We are here to support freedom. We will not be beaten.'). The media coverage in France was of the similar dimensions as the one of the march. 	A series of measures, including the prosecution of citizens who propagate terrorism on
the internet and the authorization of intelligence services to gather on-line, phone and
traveller data has been introduced for people suspected of threat.

In the days after, some number of people started questioning the real conspiracy behind the 2015 attack, carried out by the perpetrators, two French Muslim brothers of Algerian descent, who were killed in the manhunt after the attack,  presuming that they have evidences that the current French government is behind it. 

\subsection{Background information about conspiratorial rumors and mass opinion}
Conspiracy theories as political explanations have long been a part of French political culture, as well as of many others throughout the history. They are becoming particularly powerful when they are widely accepted. An empirical analysis has been reported in  \cite{Wood2015}, which concerns the nature of their support in the mass public, confirms that the chance that a person will endorse conspiracy theory can be predicted by an attraction to believes that political events are the consequence of a contest between good people and malevolent people, rather than between self-interested actors with different perspectives and priorities. They also define conspiracism as a particular form of public opinion which is distinguished from conventional opinion by its nature of animating political narratives and the latent predispositions that it activates. Of course, the question remains about what type of predispositions these conspiracy narratives will activate. Most studies of mass opinion focus on conventional sources of opinion difference, such as ideology or racial identity; some argue that conspiracism is defining the nature of the political right \cite{Sunstein2009}, and some argue that conspiracism originate in psychological predispositions \cite{Oliver2014}. 

 Accordingly, it could have been expected that the supporters of terrorism would use mass media after the attack to continue the spread of confusion and fear incited by the attack in order to achieve its goals. One of the tools in this order is indeed spreading of false information with regard to the attack. So in the days after the attack the titles appeared on the internet such as, for example "Could the January 7 Charlie Hebdo attack have been a secret service operation, or perhaps an anti-Muslim plot?", some of them even suggesting that the French president was involved in the preparation of the attack, with the intention to create negativity of the public towards Muslim population. These scenarios went so wild that to a rational person it might seem impossible that anyone could endorse them. However, the results of the public opinion poll [9] have shown that in the weeks after the attack a significant number of French citizens held conspiratorial beliefs about it (17 \%). This report gave us the motivation for the further analysis, presented in this paper.

We have checked how often 'Charlie Hebdo complot' appeared as a search query in the period of our interest (that is, the first 3 months after the attack), using Google Trends. The results of the check-up are shown in Figure 1. This quick check-up already indicated that the interest quickly faded away. The histogram shows that the number of the searches was high only in the first and second week after the attack. In the second week the interest was highest and therefore the histogram is normalized such that this week searches correspond to 100 \% or the maximal interest. Immediately after the peak interest, during the third week shown in the histogram, the poll \cite{bib4} was performed (January 20-22). Since then, the interest for this conspiracy on Google diminished.
 
\begin{figure}
  \includegraphics[scale=1.0]{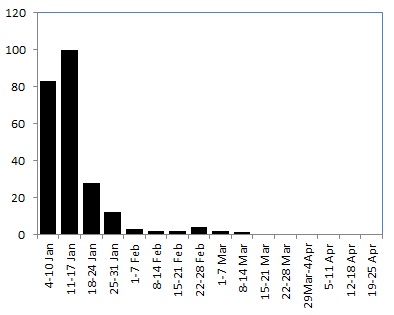}
\centering
\caption{The histogram shows Web Search interest for the phrase 'Charlie Hebdo complot'. Data Source: Google Trends (www.google.com/trends).
}
\label{fig:GoogleTrends}      
\end{figure}

\subsection{Outline of the paper}

The results of the public opinion poll \cite{bib4} have shown that in the weeks after the attack a significant number of French citizens held conspiratorial beliefs about it $(17 \%)$.  This report has been a motivation for the further analysis, presented in this paper. In Section \ref{sec:2} we discuss the data that is used for the analysis, in Section \ref{sec:3} we discuss the observations and the measurements that we have performed on the data. Our viewpoint on the presented observations is elaborated in Section \ref{sec:4}.

\section{Le Monde articles about Charlie Hebdo and on-line reactions}
\label{sec:2}
We collected 990 on-line articles mentioning Charlie Hebdo from Le Monde (one of the leading French news agencies) web site. Those articles are associated with 16490 comments posted on-line, as a reaction of the readers to the above articles. The comments are often forming short discussion threads between the readers. We identified 4825 different pseudo-names of the commentators, which means that each of the commentators reacted on average $ \sim 3.4 $ times. The average number of comments per article is $16.66$.

\begin{figure}
  \includegraphics[scale=0.4]{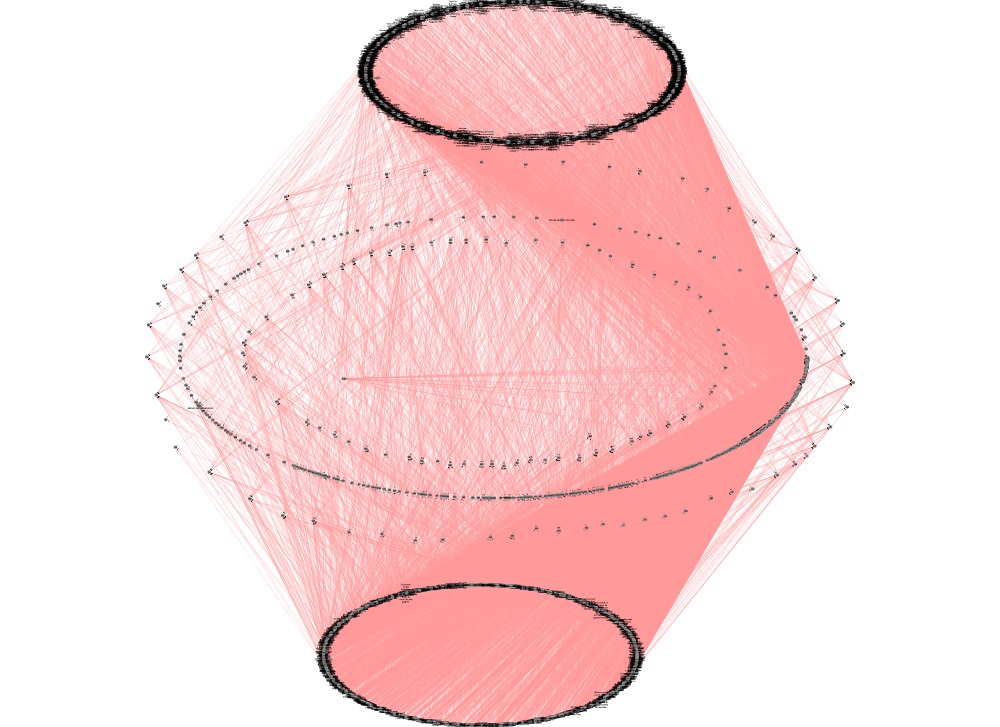}
\centering
\caption{The on-line articles published on www.LeMonde.fr  web site containing the term `Charlie Hebdo' are presented as nodes in the middle layer, while the commentators/individuals are presented as nodes in the top and the bottom layers. The links between the commentators and the articles are showing a relatively high response of the readers to the Le Monde articles about Charlie Hebdo. 
}
\label{fig:Fig1}      
\end{figure}

A network representation of all the comments and all the commentators is shown in Figure \ref{fig:Fig1}. The on-line articles are presented as 990 black nodes in the middle layer of the network, while the commentators are presented as 4825 nodes in the top and the bottom layers.  The density of links between the commentators and the articles are showing that the various response of the readers to different Le Monde articles about Charlie Hebdo. 

\begin{figure}
\includegraphics[scale=0.38]{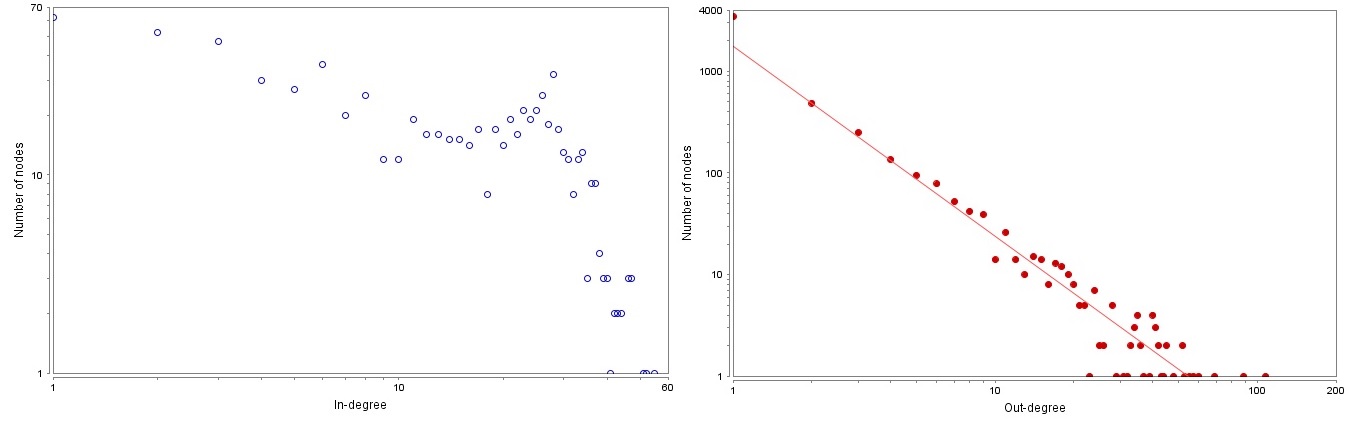}
\caption{(a) The left panel shows how many comments were posted per article. The distribution is random; most of the articles have only a very few comments, a few of them have many comments and there is an unsuual peak around 30 comments. (b) The right panel shows the activity of the commentators, fitted by a power law line.  }
\label{fig:degree}      
\end{figure}
 
In order to better understand the level and the mode of the commentators activity, we made a statistical distribution, which is based on their identities as given by the pseudo-names they use (in reality, it might happen that the same person is behind 2 or more pseudo-names, which we cannot detect). This is shown in Figure \ref{fig:degree}. The left panel shows how many comments were posted per article. The IN-degree count is irregular. The frequency of the articles with an IN-degree smaller than the average ($\sim 17$) has an decaying trend, but the statistics of the articles with a larger IN-degree is very noisy. It shows a peak at IN-degree$\sim 30$ and then it decays sharply. There is no regular statistics here.
Figure \ref{fig:degree} shows the OUT-degree of the commentators (number of comments per pseudo-name). Their activity is distributed according to a power law. The red fitting line follows a power law $y \sim a x^{-b}$ with the slope $b=1.865$, and a correlation coefficient of $0.99$.

\subsection{Selection of articles and comments mentioning Charlie Hebdo conspiracy}
\label{subsec:21}

From the above dataset containing 990 articles, we looked at the ones that contained words related with conspiracy (in French: `complot', `conspiration' or ` conjuration'). We found that there was only 9 such articles, which are listed in Appendix, in French. The bare fact that less then 1 \% of the articles mentions Charlie Hebdo in the context of conspiracy is already quite informative and shows that there is no viral contagion of conspiracy theory among Le Monde readers during this period. The number of comments per article is given in the last column of Table 1. 

We list here briefly the main subjects of the 9 articles. 

The first three articles are illustrating the conspiratorial outings of `negationists'. The main goal of negationists is denial of the historically established crimes \cite{Negationists}, and they would use conspiracy theories to support their goal. 

The first conspiracy related article that appeared after the terrorist attack was on January 15 reports an incident in Lille, when a person has been sanctioned for refusing to participate in the minute of silence and explaining that the Charlie Hebdo attack did not really happen but is a result of a conspiracy theory.
The second article writes about protective measures against conspiratorial and hateful remarks allegedly made by a young woman. This article produced a lot of reactions (56) and therefore will be analyzed in the following section.
The third articles about  a their group friends who felt offended by the cartoons about the prophet, but who could not accept that the Muslims would kill and therefore thought that there was another conspiracy behind the Charlie Hebdo attack. The claimed that killers were mercenaries hired by the secret services such as Mossad, who did it.

The second `wave' of conspiratorial beliefs comes from  Jean-Marie Le Pen, the honorary president of the ultra right winged party Front National.
Therefore, in the fourth article, Friday, January 16, in an interview with the Russian tabloid Komsomolskaya Pravda, Jean-Marie Le Pen gave controversial and undefined statements mentioning a conspiracy theory with regards to the Charlie Hebdo attack. Then, in the fifth article, he reacted to the criticism of the Vice President of his party, who suggested that Jean-Marie Le Pen had been drinking before mentioning a conspiracy theory in the attack at "Charlie Hebdo", in the earlier mentioned article. In the sixth article the same case is being discussed. All three of those articles had more then an average number of responses and they will be analyzed too in the following section.

Finally, the third `wave' of discussions about Charlie Hebdo conspiracy has a very different view point, analyzing the causes, the possible control and the prevention of those. In the seventh article (or the first one of the last set) the following question is asked: `How to distinguish right from wrong?', and a plan of the government to give more resources to the national education system to better train the youth is announced. In the following article the case of two teachers who decided to dissect in class conspiracy theories is reported. Finally in the ninth article, a summary of the earlier discussed conspiratorial ideas is made (an example: during the mobilizations for Charlie Hebdo march of January 11, the route would thus, according to conspiracists, have taken the shape of the map of Israel! etc.). These are then rationally discussed and compared. The article has a lot of reactions and they will be discussed in the following section too.

\begin{figure}
  \includegraphics[scale=0.6]{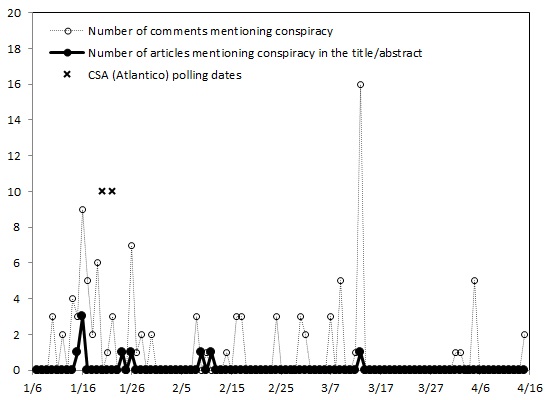}
\centering
\caption{Number of articles containing one of the words related with conspiracy (`complot', `conspiration', `conjuration') per day (black/filled circles) and the number of comments (not necessarily to those articles) containing one of the key-words. 
}
\label{fig:Fig2}      
\end{figure}

Figure \ref{fig:Fig2} shows these nine articles that are mentioning conspiracy (black circles), but also, it shows a selection of comments that are mentioning conspiracy to any of the previously  selected 990 `Charlie Hebdo' articles (white articles). In this way we can see the spontaneous discussions about Charlie Hebdo conspiracy. We see from the figure that that spontaneous activity is very low and the most of the comments have been posted as reaction to one of the articles about conspiracy. 

\section{Interpretation of the empirical observations}
\label{sec:3}

In the previous section, we described the media coverage by Le Monde, and the readers response, during the first 3 months after the attack. We identified some very minor attempts to launch conspiratorial rumors, being unsuccessful to 'go viral' on Le Monde website and during the period of observation.

Our observations need to be compared with the results of the poll \cite{bib4}, which was performed on January 20 and 22, 2015 by the Institute CSA, which is a French enterprise specialized in the studies of public opinion (these dates are also noted in Figure 4 with the 'x' symbol). Questioning was performed through the internet and the selected sample of 1000 persons was composed following the method of quote on the criteria of age, gender, profession and geographical location. The poll has been performed for the French news agency Atlantico, which published the poll results on January 25, 2015 \cite{Atlantico}. The title of their news article "17 \% des Fran\c{c}ais  pensent  que  les  attentats  de  Charlie  Hebdo  et  de la porte de Vincennes rel\`{e}veraient d'un complot: un thermom\`{e}tre de l'\'{e}tat de la soci\'{e}t\'{e}" reveals that that the population of 17 \% of the French believes that there was a conspiracy with regard to the Charlie Hebdo attack. They further also explain that this number is actually composed out of two sub-populations: 4\% of respondents are convinced that there was a conspiracy, while 13 \% lean towards this hypothesis without any conviction. The breakage of the poll results across different categories shows further details.

Our observations can also be compared with Google Trend results shown in Figure \ref{fig:GoogleTrends}. Similar to the Google Trend measurement, our measurement in Figure \ref{fig:Fig2} showed a decay in the readers' interest (please note that our data is aggregated per day and Google Trend data per week, which explains why their peak in the second week is significantly higher). It is remarkable that both Google Trend data and the data from Le Monde commentaries show an increased interest from the readers / public in the end of February 2015, which is not caused by any of Le Monde articles about conspiracy.

The great majority of the persons over 50 years (88\%) do not have any conspiratorial doubts. The situation is similar with the professionals with a higher level of education and professional responsibilities,  as 90\% of them do not have  any  doubts,  only 1  \% of is convinced about conspiracy, while 9 \% lean towards this hypothesis without any conviction. The responses collected from youth and lower level 'workers' prove to be more extreme. For example, 8\% of interviewed 18-24 year aged and workers are convinced that this is a "conspiracy" (which is double of the average).

Such interpretation of the results indeed corresponds well with our findings. Namely, the readers of Le Monde indeed are often perceived as educated professionals thus, it is not surprising that Le Monde published only very sporadically the conspiratorial rumours. However, the more-then-averaged interest of the readers for the majority of those articles shows that even though the beliefs are not present, there is certain interest for the subject.

Obviously, the situation that we observed is very different then the situation that happened after the September 11 attack, when Thierry Meyssan's book \cite{Meyssan} about 9/11 conspiracies incited an avalanche of discussions and held attention of the French citizens for quite a while. It is also true that some time is needed to publish such books (for example, a book entitled ``We Are NOT Charlie Hebdo!: Free Thinkers Question the French'' \cite{Berrett} appeared in April 2015 and it is theoretically possible that other similar books will yet appear). 

\subsection{Semantic analysis of the articles about conspiracies and the commentary}
\label{subsec:semantics}

In this section we will compare the semantic content of the articles introduced in Section \ref{subsec:21} and their commentaries.
We will analyse them in three groups: the first three articles about the disobedience and the spreading of conspiratorial beliefs in schools and public institutions, then the three articles about Jean-Marie Le Pen and his conspiracy theory and finally we will analyse the ninth article, which is more general and speak about different aspects of conspiratorial beliefs.

\begin{figure}
                \includegraphics[scale=0.3]{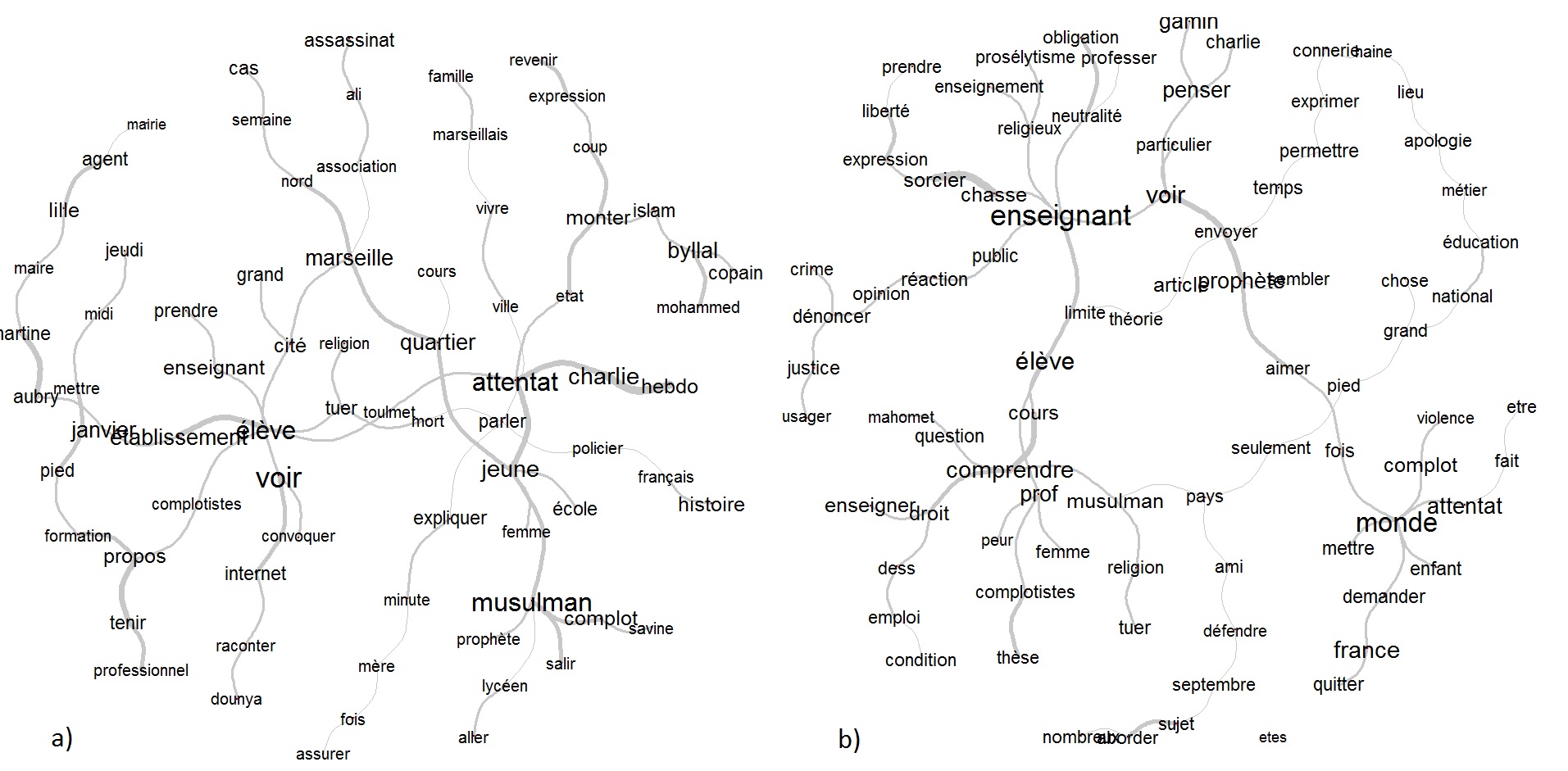}
        \caption{(a) The left panel shows semantic similarity analysis of three articles about some early conspirational statements and disrecpect for the vicitms of the attack in schools and public institutions.  (b) The comments to the articles from the left panel.  }
\label{fig:Obeissance}      
\end{figure}

The first group of articles is shown in Figure \ref{fig:Obeissance}. The semantic clouds are constructed based on the co-occurrences of the words in different parts of the text corpus (shown by the links between words) and they reveal the context in which the most frequent subjects (larger words) are used. They have been constructed using the software Iramuteq \cite{bib14}. Comparing Figures \ref{fig:ObeissanceMain} and  \ref{fig:ObeissanceComments}, we see that the contant is very similar.

\begin{figure}
                \includegraphics[scale=0.3]{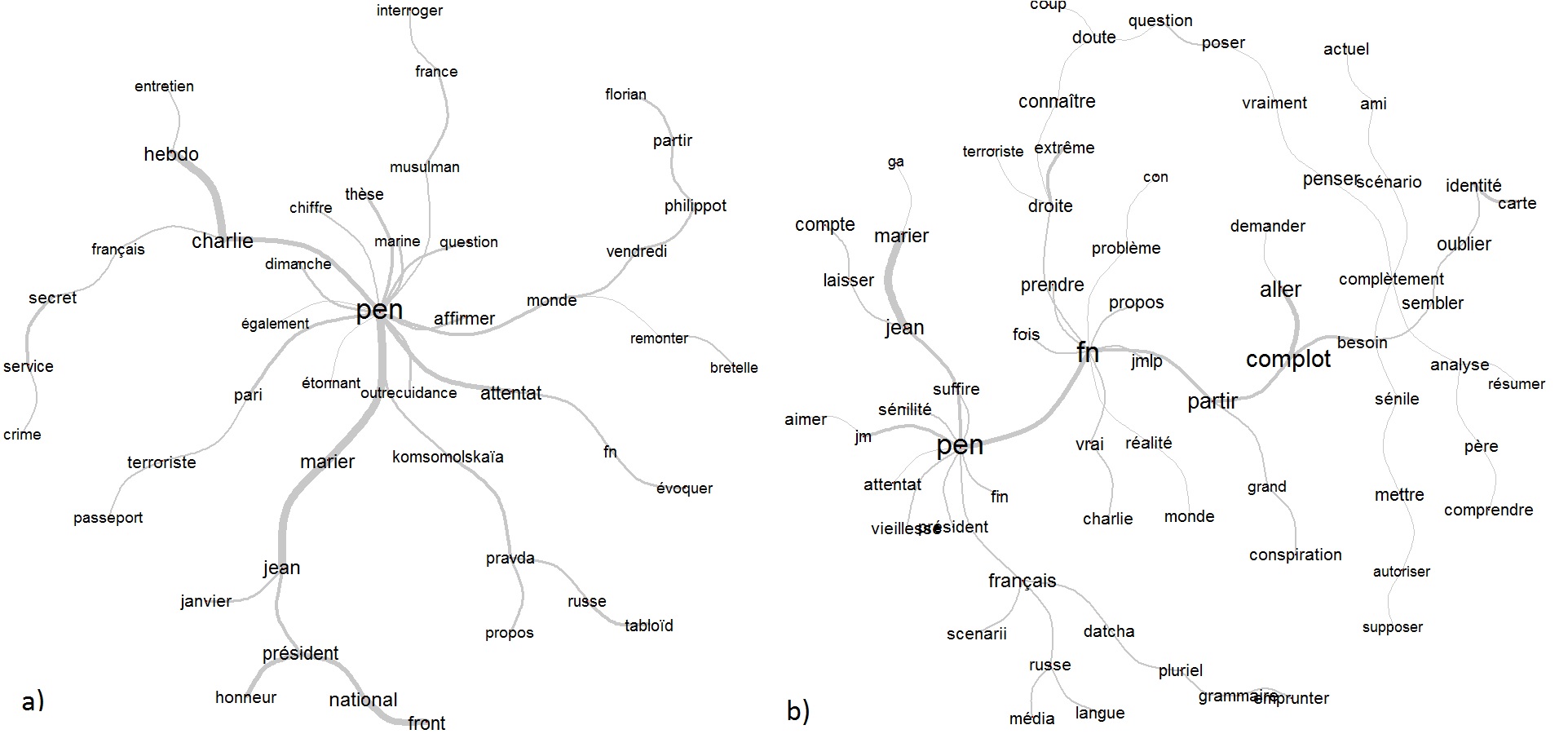}
        \caption{(a) The left panel shows semantic similarity analysis of the main text about Jean-Marie Le Pen conspirational statements.  (b) The readers' commentary to the articles (a), showing that the discussion developed rather in criticizing him then developing the conspiratorial ideas.  }
\label{fig:LePen}      
\end{figure}

\begin{figure}
                \includegraphics[scale=0.3]{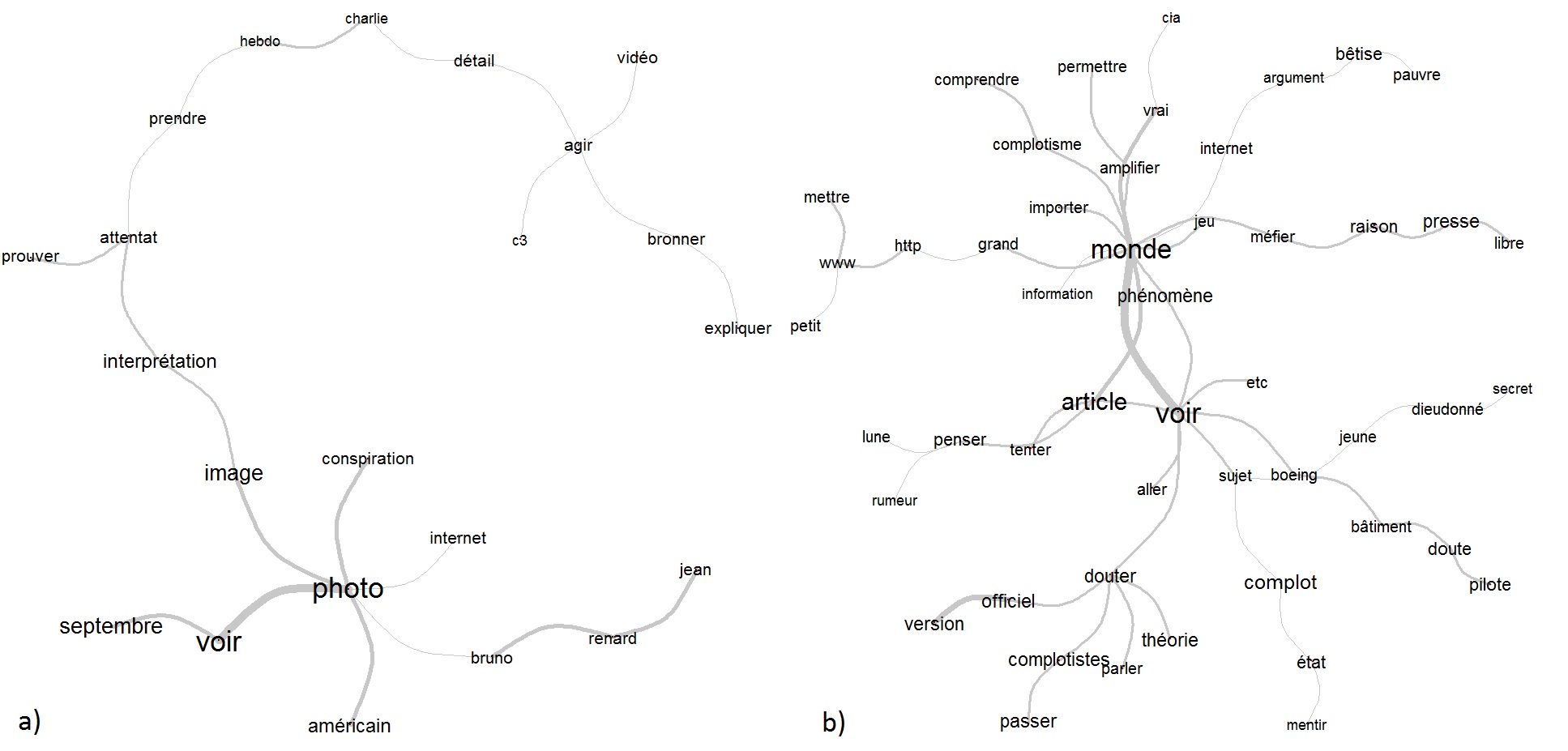}
        \caption{(a) The left panel shows semantic similarity analysis of the main text of the article ``'La conjuration des cr\'{e}dules''. (b) The comments to the article bring new content, more abstract, discussing the spreading of information through the internet, the articles, etc.  }
\label{fig:120}      
\end{figure}

The second  group of articles, about Jean-Marie Le Pen, is shown in Figure \ref{fig:LePen}.  Along with the conspiratorial version of the attack (bringing out the fact that the identity papers of the perpetrator being  left in the car in support of conspiratorists), Le Pen has also made statements on the issue of immigration. In spite of the fact  that only 8 \% of the French population declares as Muslims, he spoke of Muslims in terms of "flooding", "the invasion of our country." Such statements can not be found back in the commentary. The comments rather bring his statements in the context of the party (FN) where was at the time an honorary president.

Figure \ref{fig:120} illustrates the newest of the nine selected Le Monde articles, which reflects on those early conspiratorial attempts from a rational perspective, showed that the readers have more interest in understanding the possible causes for the onset of conspiratorial beliefs then to delve into the arguments that the conspiracists previously brought up to the public.  The left panel shows semantic similarity analysis of the main text. The main link is between the words related to the `arguments' of conpisracists, such as detail, photo, C3 (type of the car used in the Charlie Hebdo attack). In the right panel, the commentary is shown. It brings out a new content, far more abstract then the evidences of the terroristic event, discussing the spreading of information through the internet, the articles, etc. 

\section{Conclusion}
\label{sec:4}
Among other goals of violent attacks in the context of political conflict, one is to affect a broader audience beyond the physically targeted victims. In fact the wide number of people leaving Europe to join the Islamic State acts in the Middle East show that these methods work. In order to understand why and how they work, we have to be able to stage a research program that will cover this dynamics from its very incipience in the people's minds to the actual support and participation in acts of terror.
Technologically speaking this is not a totally impossible task since for business reasons the technology to follow a person activity based on its public domain presence exists and is even legal in many aspects.
Thus a systematic study of the arguments but also vocabulary and affective elements in people's interventions in internet debates might help construct the profile of the groups likely to join the trajectory between ideological sympathies to actual crimes.
The present paper is one of the emergent efforts by interdisciplinary scientists to put the basis of such a scientific program that surely will have to benefit from a very wide range of disciplines starting from psychology and linguistics and ending with computer science, statistical mechanics and percolation phase transitions. Academically this challenge is not less compelling as it connects to deep human mechanisms that can be better understood when acting in extreme and passionate contexts \cite{Vasilis}.

\section*{Acknowledgments }
This work has been performed under the grant DGA-2012 60 0013 00470 75 0. Serge Galam helped by giving many fruitful comments.
 The data analysed in the article is publicly available, and it has been kindly provided in the format appropriate for a further analysis to the author by MediaLab at SciencesPo, Paris, France.


\section*{Appendix A}
\label{sec:appendixA}
In Table 1 the selection of the Le Monde on-line articles which have been analyzed and discussed in the paper is listed. The list is a result of searching by the criterion `Charlie Hebdo' in combination with one of the following words: `complot', `conspiration' and `conjuration', which are French words for conspiracy through all the comments posted on the Le Monde articles. In the second column the date of the publication of the article is given (only day and month, as all articles were published in 2015). In the last column of the table is the number of the readers comments to the given article.

\begin{table}[!h]
\caption{Le Monde articles used in this paper.}
\label{table}
\begin{tabular}{|p{3.2cm}| p{0.6cm} | p{9cm} |p{0.4cm} |}
\hline
Title	& Date	& Abstract&	\# \\ \hline
La conjuration des cr\'{e}dules	& 13/3& {\small Pendant les mobilisations pour Charlie Hebdo, le parcours de la marche du 11 janvier aurait ainsi pris la forme de la carte d'Isra\H{e}l! `` Pour beaucoup de ces `chercheurs' du Net, il s'agit de montrer qu''' on ne se fait pas avoir ``par les m\'{e}dias.'' Ils s'imaginent plus intelligents...} & 40 \\ \hline
Au lyc\'{e}e, un cours pour d\'{e}monter les th\'{e}ories du complot	& 11/2& {\small Confront\'{e}s aux doutes de leurs \'{e}l\`{e}ves sur la r\'{e}alit\'{e} des attentats du 7 janvier contre l'hebdomadaire Charlie Hebdo, deux enseignants de r\'{e}gion parisienne ont d\'{e}cid\'{e} de d\'{e}cortiquer en classe les th\'{e}ories du complot."On a eu des confusions \'{a} cause d'Internet qui transmettait de...} & 2 \\ \hline
Attentats de janvier : les lyc\'{e}ens en proie aux th\'{e}ories du complot & 	9/2& {\small Depuis les attentats contre "Charlie Hebdo" et l'HyperCasher de la porte de Vincennes, \'{a} Paris, les th\"{e}ses complotistes circulent sur Internet. Comment distinguer le vrai du faux ?... Le gouvernement a annonc\'{e} vouloir donner davantage de moyens \'{a} l'\'{e}ducation nationale pour mieux former...} & 0 \\ \hline
Attentats de Paris: Jean-Marie le Pen d\'{e}fend la th\`{e}se du complot	& 26/1& {\small Le Pen avait "peut-\^{e}tre pris un peu de vodka" avant de donner un entretien au tablo\''{i}d russe Komsomolska\''{i}a Pravda, dans lequel il laissait entendre ses doutes \'{a} propos de l'attentat \'{a} Charlie Hebdo.. "Outrecuidance", avait r\'{e}pondu M. Le Pen. Mais \'{a} l'occasion de cette galette des...}	& 25 \\ \hline
Le ton monte au FN entre Jean-Marie Le Pen et Florian Philippot	& 24/1& {\small Le pr\'{e}sident d'honneur du parti a fustig\'{e} samedi l'"outrecuidance" du vice-pr\'{e}sident, qui a sugg\'{e}r\'{e} que Jean-Marie Le Pen avait bu avant d'\'{e}voquer une th\`{e}se conspirationniste lors de l'attentat \'{a} "Charlie Hebdo"... L'attentat de Charlie Hebdo ? "Cela ressemble \'{a} une...} & 27 \\ \hline
Charlie Hebdo: la petite musique conspirationniste de Jean-Marie Le Pen	& 16/1& {\small Apr\`{e}s la parution, vendredi 16 janvier, dans le tablo\''{i}d russe Komsomolska\''{i}a Pravda d'une interview choc, Jean-Marie Le Pen maintient aupr\`{e}s du Monde la quasi-totalit\'{e} de ses propos publi\'{e}s concernant la tuerie de Charlie Hebdo. Dans cet entretien, le pr\'{e}sident d'honneur du Front...} & 30 \\ \hline
Les attentats vus par des jeunes marseillais : "C'est un complot pour salir les musulmans"	& 16/1& {\small Byllal, 24 ans, Mohammed, 22 ans, et leurs deux amis, \'{a} peine plus jeunes, ne d\'{e}filent pas pour Charlie Hebdo. Ils sont venus faire du shopping. "On pense que toute cette histoire est une manipulation, attaque Byllal. Ces tueurs \'{e}taient des mercenaires engag\'{e}s par les services secrets...} & 17 \\ \hline
Bobigny: une enseignante mise \'{a} pied pour avoir tenu des propos complotistes	& 16/1& {\small Une mesure conservatoire, "en accord avec la pr\'{e}fecture de la Seine-Saint-Denis", a pr\'{e}cis\'{e} au Monde son pr\'{e}sident, Patrick Toulmet, et qui fait suite \'{a} des propos complotistes et haineux qu'aurait tenus la jeune femme \'{a} des \'{e}l\`{e}ves de baccalaur\'{e}at professionnel, lundi 12 janvier, au sujet des deux...	} & 56 \\ \hline
A Lille, un agent municipal sanctionn\'{e} pour apologie du terrorisme	& 15/1& {\small  Ce dernier a refus\'{e} de participer \'{a} la minute de silence jeudi dernier et aurait expliqu\'{e} que l'attentat contre la r\'{e}daction de Charlie Hebdo n'avait pas eu lieu et que tout cela n'\'{e}tait que complot. "Le procureur a re\c{c}u une plainte de la mairie de Lille, confirme le parquet. Il a...	} & 5 \\ \hline

\end{tabular}
\end{table}
\end{document}